# Single Crystalline Colloidal Quasi-Two-Dimensional Tin Telluride

Fu Li,[1] Jiecai Fu,[2] Abderrezak Torche,[1] Sascha Kull,[1] Andreas Kornowski,[1] Rostyslav Lesyuk,[1,3,4] Gabriel Bester,[1] Christian Klinke[1,3,5,*]

*[1] Institute of Physical Chemistry, University of Hamburg, Martin-Luther-King-Platz 6, 20146 Hamburg, Germany*
*[2] Key Laboratory for Magnetism and Magnetic Materials of the Ministry of Education, School of Physical Science and Technology, Lanzhou University, 730000 Lanzhou, P. R. China*
*[3] Institute of Physics, University of Rostock, Albert-Einstein-Strasse 23, 18059 Rostock, Germany*
*[4] Pidstryhach Institute for applied problems of mechanics and mathematics of NAS of Ukraine, Naukowa str. 3b, 79060 Lviv, Ukraine & Department of Photonics, Lviv Polytechnic National University, Bandery str. 12, 79000 Lviv, Ukraine*
*[5] Department of Chemistry, Swansea University – Singleton Park, Swansea SA2 8PP, United Kingdom*

\* Corresponding author: christian.klinke@uni-rostock.de

**Abstract**

Tin telluride is a narrow gap semiconductor with promising properties for IR optical applications and topological insulators. We report a convenient colloidal synthesis of quasi-two-dimensional SnTe nanocrystals through the hot-injection method in a non-polar solvent. By introducing the halide alkane 1-bromotetradecane as well as oleic acid and trioctylphosphine, the thickness of two dimensional SnTe nanostripes can be tuned down to 30 nm, while the lateral dimensional can reach 6 $\mu$m. The obtained SnTe nanostripes are single-crystalline with a rock-salt crystal structure. The absorption spectra demonstrate pronounced absorption features in the IR range revealing the effect of quantum confinement in such structures.

Owing to the high potential for infrared and near infrared detectors, solar cells and thermoelectric applications, narrow-bandgap semiconductor nanocrystals have triggered intensive investigations in the recent years.[1-6] Tin telluride (SnTe), as a IV-VI narrow bandgap semiconductor (0.18 eV, bulk), has a rock-salt crystal structure ($Fm\bar{3}m$) with mirror symmetry in contrast to other tin chalcogenides (SnS and SnSe) with orthorhombic layered crystal structure.[7, 8] SnTe has also been considered as a remarkable topological crystalline insulator (TCI) based on its crystal symmetry, and it has a potential for mid-IR applications, as well as photovoltaic devices.[9-12] So far, attempts to obtain solution-based SnTe nanocrystals (NCs) only yielded zero-dimensional (0D) or one-dimensional (1D) nanostructures.[13-15] Here, we report a synthesis protocol for large and thin SnTe nanosheets and nanostripes with high crystallinity and the thicknesses down to 30 nm and largest lateral sizes of up to 6 $\mu$m. To the best of our knowledge, this is the first instance for 2D colloidal SnTe nanostructures. The obtained nanocrystals have been investigated by means of transmission electron microscope (TEM), high resolution TEM, powder X-ray diffraction (XRD), selected area electron diffraction pattern (SAED), atomic force microscopy (AFM), energy-dispersive X-ray (EDX) analysis and Fourier transform infrared (FTIR) measurements. The absorption spectra of nanosheets and nanostripes in ensemble showing distinct features in the mid-IR are rationalized with the support of DFT calculations.

**Results and Discussion**

As a non-flammable and reactive Sn (II) precursor, tin acetate was used together with elemental Te dissolved in trioctylphosphine. In a typical synthesis, 0.25 mmol tin (II) acetate (TA), 6.4 mmol oleic acid (OA) and 1.0 mmol trioctylphosphine (TOP) were mixed with 10 ml diphenyl ether (DPE) as solvent. After degassing the solution at 75°C for 2 h, 1-bromotetradecane (1-BTD) was added as Br source, followed by heating the system to the reaction temperature of 250°C. Following the fast injection of the Te precursor (TOP-Te), the reaction changed color rapidly from yellow to greyish yellow. After the color change, the reaction solution was cooled down to room temperature, followed by purification and storage in toluene.

The as-synthesized 2D SnTe nanocrystals (NCs) consist mainly of 2D nanostripes (NSPs), and some stick-shaped nanocrystals (SNCs) (Figure 1). As a side-product, cubic SnTe nanoparticles (NPs) can be observed as well. A typical unit cell of the cubic phase of β-SnTe is shown in Figure 1a.[16, 17] Figure 1b-e shows typical TEM images, SAED, and an AFM image of the SnTe NSPs prepared at a reaction time of 250°C for 1.5 min. The stripes can range from 1.5 $\mu$m to 5.8 $\mu$m in length with a mean length of 3.2 $\mu$m derived from the scanning electron microscope (SEM) images (Figure 1f). The SAED patterns shown in Figure 1d display well-defined dot patterns representing the planes of (0$\bar{2}\bar{2}$), (0$\bar{2}$0), (0$\bar{2}$2), (002), (022), (020) and (02$\bar{2}$) of a single SnTe NSP (zone axis <100>). This demonstrates that the obtained SnTe NSPs are single-crystalline and single-phase structures. The AFM image and the corresponding height analysis show that the SnTe NSP has a flat surface morphology and the thickness of approximately 30 nm (Figure 1e). Powder X-ray diffraction (Figure 1g) for the sample collected in a capillary demonstrates that all the diffraction peaks can be indexed as cubic crystal structure of SnTe (*Fm$\bar{3}$m*, 225, 00-046-1210), while the XRD pattern of the drop-casted sample exhibits pronounced reflexes of (200) and (400) due to the texture effect. This implies that the NSPs lay perpendicular to the [100] direction, which is defined here as the thickness direction. In addition, the measured angles between the tip-edges are 135° and between a tip-edge and a side-edge 90°, as will be explain later in the text (Figure 1c).

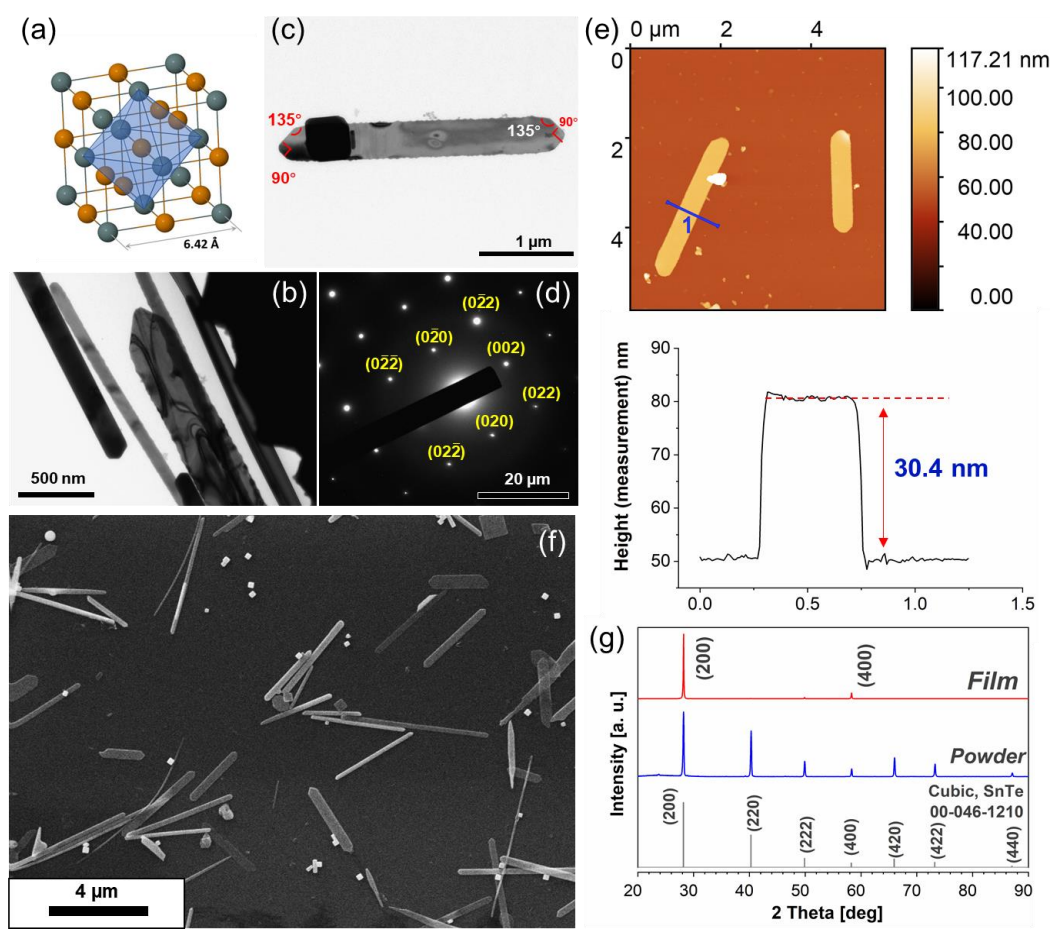

**Figure 1.** (a) Face-centered cubic unit cell (rock salt structure) of β-SnTe (grey: Sn, orange: Te). Each cation is coordinated by six anions forming a coordination octahedron. (b, c) TEM image of SnTe NSPs using the standard synthesis process. (d) SAED pattern of a SnTe single NSP. (e) AFM image and height profile of a SnTe NSP. (f) SEM image of SnTe NSPs, NPs and NRs. (g) XRD patterns of drop-casted sample and sample in capillary tube of SnTe NCs.

High resolution TEM (HRTEM) characterizations of SnTe NSPs and SnTe NPs, together with the corresponding fast Fourier transform (FFT) analysis were performed as shown in Figure 2, displaying the crystallography, exposed facets and the growth direction of SnTe NSPs (main product) and NPs (byproduct). The measured lattice spacing of approximately 0.32 nm, which are perpendicular, confirms the two perpendicular in-plane facets (002) and (020) belong to the {100} family (Figure 2b, c). With the already-known angles from Figure 1c and the crystal orientation, the right-angled edges of the NSP can be determined to be parallel to the (010) and (001), as well as their parallel facets ($0\bar{1}0$) and ($00\bar{1}$). Based on FFT analysis, we deduct that the long axis points along the [011] or [$0\bar{1}\bar{1}$] directions and the analysis of according HRTEM images yields a lattice spacing of 0.227 nm (Figure 2b, c, f, g). An energy-dispersive X-ray (EDX or EDS) line scan analysis of the NSP (Figure S1a) and the corresponding EDX element analysis (Figure S2) yield a highly stoichiometric ratio of Sn to Te (1:1), and also a small amount of oxygen in the SnTe NSPs and NPs. EDX mapping also confirmed an even distribution of the elements tin, tellurium, and on the surface located oxygen, which is clearly recognizable as a layer around the SnTe particle (Figure S1b, c). The origin of this oxide layer could be a result of surfactants like oleic acid and/or acetate, a weak oxidation of tellurium, tin tellurate or a $SnO_X$ shell founded on SnTe nanocubes reported by Guo *et al.*[13] However, in HRTEM images we can clearly see areas with interrupted fringes appearing amorphous (Figure 2b inset). A uniform distribution of a small amount of oxygen through the nanostripe was also found in the line-scan profile in Figure S1, indicating the presence of oxygen on the SnTe nanostripes as well. Based on the EDX mapping data, the sizes of the mapping images of element Sn and O are slightly larger than that of element Te, implying that the phenomenon of a $SnO_X$ shell on SnTe NCs. This shell might protect the synthesized SnTe nanocrystals from being oxidized further, giving the nanocrystals more stability, which is already demonstrated by Guo *et al.*[13]

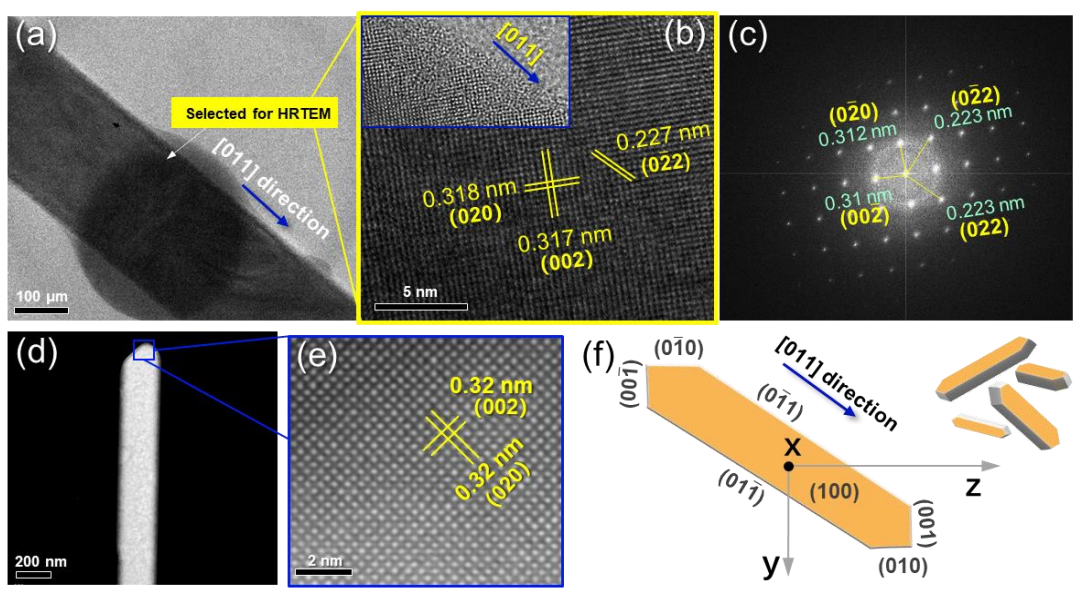

**Figure 2**. HRTEM characterizations of SnTe NSPs and the corresponding FFT analysis (a-c). (d, e) STEM characterization of SnTe nanoparticles (NPs). (f, g) STEM characterization of the right-angled edges of the SnTe NSP. (h) Schematic representative of SnS nanostripes.

The morphology evolution was investigated by taking aliquots during the reaction followed by TEM and SAED analysis (Figure S3). At the early stages, before 30 s, there are no defined nanostructures with cubic phase observed, only $SnO_2$ nanostructures are found as confirmed by SAED and TEM characterization (Figure S4). The SAED pattern indicates the $SnO_2$ nanostructures align already during the first stages of the synthesis, which may hint to oriented attachment as growth mechanism of the earlier-formed $SnO_2$ particles.[18, 19] Afterwards, the SnTe NSPs with defined shape (approximately 2 μm length, 100 to 180 nm width) form within the first minute of the reaction (Figure S3c). As the reaction proceeds, the NSPs grow longer and the widths become larger. The largest of them reach 3.8 μm and 1.2 μm in length and in width respectively after 3.5 min of reaction (Figure S3e). In addition, a small fraction of particles was also formed with the same cubic phase. The $SnO_2$ product is assumed to origin from the introduction of a tiny amount of oxygen when when injecting 1-BTD during the reaction (5 s after hot-injection). Its formation is assumed to be independent from the nucleation and growth of SnTe nanocrystals afterwards (1 min after hot-injection). The TEM analyses provide a possible clue for the formation of SnTe NSPs. The fast nucleation occurs after the hot-injection of the tellurium precursor, followed by the anisotropic growth process based on different growth speed on different facets, finally forming the defined morphology.[20, 21]

As promising symmetry-breaking ligands for colloidal nanocrystal syntheses, halide ions have some advantages such as relatively high electronegativities and

reactivities.[22-24],[25] For comparison, reference syntheses without 1-BTD were also performed (Figure S5). However, there are only agglomerates produced if no 1-BTD was used during the synthesis. We found that the amount of 1-bromotetradecane (1-BTD) is one of the key factors for obtaining relative uniform and well-defined 2D SnTe NSPs. The amount of 1-BTD as structure giving factor was studied to investigate the influence of the halide ligand on the synthesis and to find the optimal amount to obtain 2D NSPs (Figure 3). It demonstrates that irregular-shaped platelets are formed together with long SNCs and NPs instead of well-defined NSPs, when the 1-BTD amount is less than 0.64 mmol (Figure 3a, b). When the amount reaches 0.64 mmol, most of the product is composed of NSPs together with a few NRs and cubic NPs (Figure 3c). This indicates that small amounts of 1-BTD together with OA and TOP as ligands will passivate the {100} facets, impelling the nuclei to favor the formation of NSPs with right-angled edges and small height compared to the large lateral size (Figure 2f, h). Larger amounts of 1-BTD lead to a larger number of long rods (or wires) (Figure 3d). This may result from the replacement of other ligands by stronger X-type ligand ($Br^-$) and destabilization of the (100) facets and faster growth along the [011] direction. The corresponding XRD diffractograms show strong texture effect (Figure 3e).

AFM measurements show that the NSPs synthesized with 0.34 mmol of 1-BTD possess similar thicknesses (around 30 nm, Figure 3g) as the ones with 0.64 mmol 1-BTD, which is the standard synthesis (Figure 1e), while long stick-shaped nanocrystals are synthesized with 0.96 mmol 1-BTD (thickness around 80 nm, Figure 3h). These long sticks possess similar diameters as the ones obtained in the standard synthesis (approximately 90 nm in thickness, Figure S6).

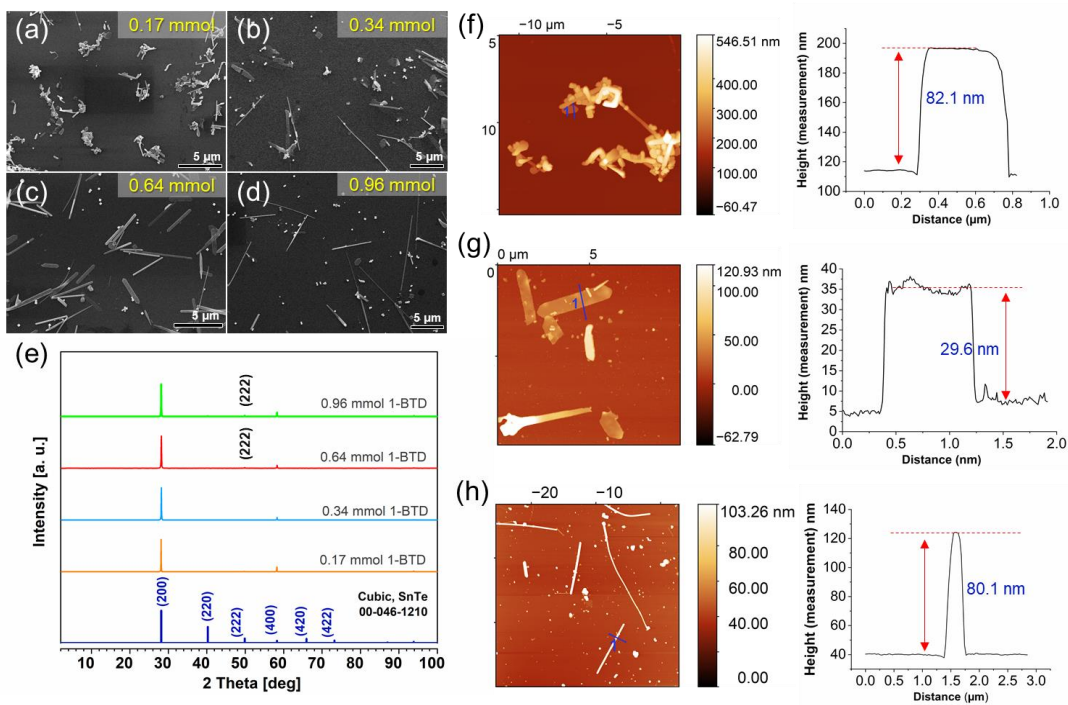

**Figure 3**. Synthesized SnTe nanostructures prepared with different amounts of 1-BTD: (a) 0.17 mmol, (b) 0.34 mmol, (c) 0.64 mmol, (d) 0.96 mmol. (e) XRD diffractograms of the samples in a-d. (f-h) AFM images and height measurements of the synthesized SnTe nanostructures from Figure 3a, b, d.

In contrast to the synthesis with the assistance of 1-BTD, the reaction in the presence of 1-CTD leads to the formation of NSs beside NSPs. We found that 1-chlorotetradecane (1-CTD) could also assist to form 2D SnTe nanostructures (Figure S7). The reaction temperature was found to be optimal in the range of 220–260°C for NSs/NSPs formation. When the synthesis temperature was lowered to 220°C in the case of BTD the nanostripe shape still dominates in the product (Figure S9). By varying the reaction temperature between 220°C and 260°C, in the CTD assisted synthesis we can observe a tendency more clearly. The 2D nanostructure shape evolves from thick squares or irregular shapes (112.7 nm, Figure S7a), to thick elongated stripe-like shapes (99.1 nm, Figure S7b). All the XRD diffractograms for these samples display pronounced (200) and (400) reflexes representing strong texture effect (Figure S8). The reason for obtaining the product synthesized by using chloride alkanes can be explained by the nucleophilicity difference of Cl$^-$ and Br$^-$ ions (Cl$^-$ < Br$^-$). The slower ion release will occur with Cl$^-$ compared to Br$^-$, based on their ion sizes to produce lower amounts of ions available at the early stages of the reaction, in particular at lower temperature (220°C). It will then result in a less dense coverage on each facet, which leads to a smaller degree of growth anisotropy for the nanocrystals. Hence, the synthesis with 1-CTD shows a lower yield of 2D morphology with more 3D NPs instead (Figure S7).

As has been found earlier, halogen alkanes units and their corresponding released halide ions can be both found on the surface of 2D nanoparticles when halide alkanes are adopted as ligands in the reaction.[26] In order to decouple the possible effect of two species in our synthesis, we introduced the halogens in the complex with tin precursors using the tin chloride precursor instead of tin acetate. It showed that 2D SnTe nanostructures can be also obtained (Figure S10). This supports the argument about the key role of the halogen ions in breaking the symmetry during the nucleation and growth of the 2D SnTe NCs. More than half of the nanocrystals are etched after collection from the synthesis, which is independent of the amount of tin precursor in the reaction. Tin (II) bromide can also lead to similar products as tin (II) chloride (Figure S11). The AFM characterization shows that the thickness is around 76 nm (Figure S12). We note a strong etching effect, which appears at the beginning of NS formation (Figure S13). The taking-aliquot experiment indicates that the etching by halide ions happens rapidly from the very beginning during the formation of NSs. As known, halide ions released from halide alkanes are the actually functioning ligands instead of the original halide alkanes themselves in IV-VIA group semiconductors.[27] By means of the specific binding preference on crystal surfaces, or the complexation with metal atoms, or the etching process, halide ions have been proven as good optional ligands for obtaining 2D nanostructures in solution.[22, 24, 28] Further, we found that for the NS formation, halide ions can be introduced from salts such as LiCl or LiBr dissolved in dimethylformamide (DMF) and added to the reaction flask. (Figure S13). However, in this case, along with NSs we observe structures with undefined or disturbed 2D morphology. Therefore, the involvement of halide ions, especially halogen alkanes is proven to be vital for the symmetry breaking in 2D cubic-phase SnTe nanocrystal synthesis. Further, we can conclude that alkane chains in the syntheses with halide alkanes play an important role in stabilizing nanocrystals during the reaction, avoiding the formation of agglomerates and avoiding the etching process.

Note, the occurrence of the etching on SnTe is assumed to be mainly due to the change of the chemical environment when the tin chloride/bromide is used.[27] The salt addition into the reaction can affect the polarity of the solvent. Further, it also alters the capping conditions of the organic ligand layers on the crystal surface, which triggers the etching process when the ligand layer conditions change, followed by a change of the solubility of SnTe. This enables the dissolution of SnTe at the very early stage of the growth after the sheets form in the case of using tin chloride/bromide or LiCl/LiBr.

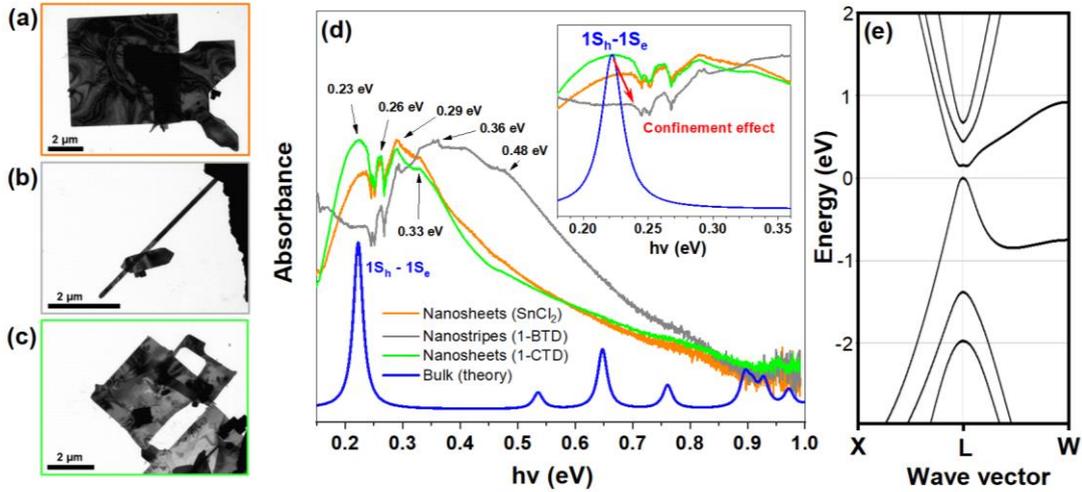

**Figure 4.** Optical absorption spectra of SnTe nanosheets and nanostripes synthesized respectively using different halides and halide sources: $SnCl_2$ as precursor (a), 1-BTD molecules as halide ligands (b) and 1-CTD molecules as halide ligands (c). (d) Comparison between experimental spectra (orange, green, gray) for nanosheets and nanostripes with theoretical (DFT-PBE) bulk absorption (in blue) where we see no absorption peaks in the range 0 to 0.5 eV except the $1S_h$-$1S_e$ transition (the range 0.15–0.35 eV magnified in the inset). Note, the first theoretical absorption peak has been blue-shifted to match the first experimental absorption feature in nanosheets (a); (e) calculated electronic bands structure of bulk SnTe near the L-point, the Fermi level is set to 0.

The absorption spectra of SnTe nanostructures obtained with different halide sources recorded by Fourier-transform infrared (FTIR) spectroscopy are shown in Figure 4. We find pronounced features in the mid-IR range. It has been reported that oleic acid has an absorption band from 3000 nm to 4000 nm (~410-310 meV)[14], which we think is responsible for the observed peaks at 0.33, 0.36 and 0.4 eV. Both NSs and NSPs demonstrate three strong absorption peaks at 230 meV, 260 meV and 290 meV, from which the latter two features show the same shape and position in the three samples. This latter observation suggests that the two peaks at 260 meV and 290 meV are potentially structure independent as they do not show confinement effect and no shape change. The first absorption peak at around 230 meV is however blue-shifted in the 30 nm nanostripes by 20 meV as compared to the 80 nm NSs and also shows a change in the shape indicating that it is potentially a structure dependent peak. In order to get more insights about the measured spectra, we calculated the absorption spectrum of bulk SnTe using the independent-particle approximation (i.e. no excitonic effects) using fully relativistic (i.e. including spin-orbit interaction) density functional theory (DFT) calculation at the PBE level as implemented in the Quantum-espresso package.[29]

Unfortunately, the *ab-initio* simulation of a simulated 30 nm SnTe slab is prohibitively expensive and cannot be performed. Based on the band structure (see Figure 4, e), we note that bulk SnTe is an L-point direct band gap semiconductor where in the range 0.0 to 1.0 eV only the valence band maximum (VBM or $1S_h$) and the conduction band minimum (CBM or $1S_e$), CBM+1, CBM+2 contribute to the optical spectrum. Moreover, in contrast to standard Γ-point semiconductors (such as CdSe, ZnS etc.), bands with heavy-, light-hole and split-off character are absent near the Fermi level around the L point in SnTe. The theoretical optical spectrum (Figure 4, d) of bulk SnTe shows only one absorption peak in the range 0.0 to 0.5 eV, which corresponds to the fundamental $1S_h$-$1S_e$ transition in bulk SnTe. Since the Bohr radius of the bulk SnTe is 95 nm, lowering the thickness up to 80 nm and 30 nm results in a blue-shift of this fundamental transition to match the first observed peak in the 80 nm and 30 nm NSs and nanostripes respectively.[13] The absence of other absorption peaks in the range 0.0 to 0.5 eV confirms further that two peaks at 260 meV and 290 meV are potentially due to ligand, defect or surface trapped states. To verify whether they can also be due to higher excitonic transitions (which are not described in our actual calculation) as suggested by previous works,[13, 14] we later calculate the exciton Rydberg energy for SnTe. Using the effective mass approximation to describe the one-dimensional confinement in NSs and nanostripes, the change in the band gap ($E_g$) in theses nanostructures as compared to the bulk value is given by : $\Delta(L) = \pi^2 * \frac{\hbar^2}{2*m_{eff}*L^2}$, with $m_{eff} = 0.036 \cdot m_0$, and $L$ being the reduced electron-hole effective mass and the thickness of the simulated slab respectively ($m_0$ is the free electron mass). Note that the effective mass in SnTe is highly anisotropic and non-parabolic,[30] however, this should not affect the qualitative description of the optical spectra adopted here within the effective mass approximation. DFT-PBE calculations with spin-orbit yield a hole effective mass of 0.13 in the L-W direction and 0.27 in the L-X direction while for the electron it is 0.046 and 0.042 in the L-W and L-X directions respectively (atomic units are used). This gives an averaged electron-hole mass of $0.036 \cdot m_0$ and the corresponding change in the band gap is then 12 meV for nanostripes of 30 nm thickness and only 1.5 meV for NSs with 80 nm. The difference being 10 meV is in qualitative agreement with the experimental value of 20 meV, which corresponds to the blue-shift of the first absorption in nanostripes as compared to NSs. This revealing the effect of quantum confinement in thinner nanostructures.

Based on the values for effective charge carrier masses we further calculate the exciton Rydberg energy for SnTe: $R_X = \frac{\mu}{m_0 \xi_r^2} R_H$, where $\mu$ is the reduced mass calculated above, $m_0$ – electron mass, $\xi_r$ – static dielectric constant, $R_H$ – Rydberg energy of the hydrogen atom (13.6 eV). With the value of $\xi_r$=1770,[31] the $R_X$=1.64×10$^{-7}$ eV. The energy difference between the first and second excitonic transitions will be: $E_2^{exc} - E_1^{exc} = \frac{3}{4}R_X$ derived from the expression for the exciton energy $E_n^{exc} = E_g - \frac{R}{n^2}$, with n=1, 2, 3... Thus, the further observed transitions after the 1S$_h$-1S$_e$ (at 0.26, 0.29, 0.36 eV) cannot be attributed to the series of excitonic states. With such a tiny $R_X$ these transitions could be resolved neither at room temperature nor at usual cryogenic temperatures. Further, the hypothetical appearance of plasmonic resonances was suggested by Gou et al.[10] based on a self-doping effect. In our case the stoichiometry of 1:1 (Sn:Te) was found and we rather exclude the plasmonic resonances in this range of wavelengths.

**Conclusion**

We demonstrated the first colloidal synthesis approach of quasi-two-dimensional SnTe nanocrystals with narrow band gap. The introduction of halide alkanes to the synthesis of cubic-phase SnTe nanocrystals in solution might provide a means to also manipulate other 2D nanocrystal syntheses, such as for HgTe. The small thickness (down to 30 nm) and the large lateral size (6 $\mu$m) with high crystallinity, and the colloidal approach manifests a promising application of theses nanomaterials in infrared optical applications, electronic or optoelectronic devices, as well as low-dimensional topological insulators.

**Experiment details**

**Chemicals and Materials.** Oleic acid (OA, 90%), trioctylphosphine (TOP, 97%), diphenyl ether (DPE, ≥99%), 1-bromotetradecane (1-BTD, 97%), 1-chlorotetradecane (1-CTD, 98%), N, N-dimethylformamide (DMF, pharmaceutical secondary standard), lithium chloride (LiCl, anhydrous, ≥99%), tin (II) bromide (SnBr$_2$), and toluene were purchased from Sigma-Aldrich. Tin (II) acetate (Sn (CH$_3$CO$_2$)$_2$, anhydrous, ≥99.99%) and tellurium powder (Te, 99.997%) was purchased from Alfa Aesar. Tin (II) chloride (SnCl$_2$, ≥99.99%) was purchased from Thermo Fisher Scientific. All the chemicals were used as-received without additional purification. Trioctylphosphine-Te (TOP-Te) was prepared and stored in a glovebox and all the syntheses were carried out applying standard air-free Schlenk-line techniques.

**Material Characterization.** Transmission electron microscope (TEM) images and selected area electron diffraction (SAED) patterns were obtained using a TEM with the type of JEOL-1011(LaB$_6$, 100 kV). Samples for TEM measurements were prepared by dropping a 10 µL diluted toluene dispersion onto carbon-coated TEM grids. Atomic-resolved images were recorded by a probe-corrected high-angle annular dark-field scanning transmission electron microscopy (HAADF-STEM, FEI Themis Cubed 300) with a probe semi-convergence angle of 24.7 mrad and collect angle range of 70-200 mrad. To avoid beam damage, the atomic HAADF-STEM image was obtained by an image series composed of sixteen fast exposure images of the same region following by a post-acquisition image distortion correction and averaging (Smart Align).[32] The high resolution (HR) TEM images were performed on a double corrected (CETCORR and CESCORR, CEOS) JEOL 2200 FS operating at 200 kV. Scanning electron microscopy (SEM, LEO 1550) was also applied for morphology information of SnTe nanocrystals. Wide-angle powder X-ray diffraction (XRD) patterns were collected using Philips X'Pert X-ray diffractometer with Cu Kα radiation ($\lambda$ = 0.154 nm). Fourier transform infrared (FTIR) measurements were carried out by drying the nanocrystals and putting the powders on a diamond-ATR unit (Bruker VERTEX 70). The FTIR measurements are performed with a range from 400 to 8000 cm$^{-1}$.

**Synthesis of 2D SnTe nanosheets.** In a typical synthesis of SnTe NSs, Sn (CH$_3$CO$_2$)$_2$ (TA, 0.25 mmol), oleic acid (OA, 2.0 mL), trioctylphosphine (TOP, 0.5 mL) and diphenyl ether (DPE, 10 mL) were mixed and dissolved in a 50 mL three-neck flask, then dried and degassed under vacuum at 75°C for 120 min. The solution was then heated to reaction temperature 250°C after adding certain amount of 1-bromotetradecane (1-BTD), followed by the rapid injection of 1M TOP-Te precursor solution. The heating mantle was removed when the solution color changed from yellow to greyish yellow. The resultant nanostructures were then purified by centrifugation with toluene at 4000 rpm for 3 min (2-3 times). The product could then be re-suspended in toluene for further characterization or storage.

ACKNOWLEDGMENT

This work was supported by the China Scholarship Council (CSC), the ERC Starting Grant "2D-SYNETRA" (Seventh Framework Program FP7; Project: 304980) and the German Research Foundation DFG in the frame of the Heisenberg scholarship KL 1453/9-2. We thank Elaine Rahmstorf, Robert Schön and Stefan Werner (University of Hamburg) for help with the FTIR, SEM and TEM measurements.

**References**

[1] L. Sun, J. J. Choi, D. Stachnik, A. C. Bartnik, B.-R. Hyun, G. G. Malliaras, T. Hanrath, F. W. Wise, *Nat. Nanotechnol.* **2012**, 7, 369.
[2] Y. Du, B. Xu, T. Fu, M. Cai, F. Li, Y. Zhang, Q. Wang, *J. Am. Chem. Soc.* **2010**, 132, 1470.
[3] M. C. Beard, J. M. Luther, A. J. Nozik, *Nat. Nanotechnol.* **2014**, 9, 951.
[4] M. Ibáñez, R. Zamani, A. LaLonde, D. Cadavid, W. Li, A. Shavel, J. Arbiol, J. R. Morante, S. Gorsse, G. J. Snyder, *J. Am. Chem. Soc.* **2012**, 134, 4060.
[5] J. S. Liu, X. H. Li, Y. X. Guo, A. Qyyum, Z. J. Shi, T. C. Feng, Y. Zhang, C. X. Jiang, X. F. Liu, *Small* **2019**, 15, 1902811.
[6] J. Feng, X. Li, Z. Shi, C. Zheng, X. Li, D. Leng, Y. Wang, J. Liu, L. Zhu, *Adv. Optical Mater.* **2019**, 1901762.
[7] M. Safdar, Q. Wang, M. Mirza, Z. Wang, K. Xu, J. He, *Nano Lett* **2013**, 13, 5344.
[8] T. Feng, D. Zhang, X. Li, Q. Abdul, Z. Shi, J. Lu, P. Guo, Y. Zhang, J. Liu, Q. J. Wang, *ACS Appl. Nano Mater.* **2020**, 3, 674.
[9] T. H. Hsieh, H. Lin, J. Liu, W. Duan, A. Bansil, L. Fu, *Nature communications* **2012**, 3, 982.
[10] Y. Tanaka, Z. Ren, T. Sato, K. Nakayama, S. Souma, T. Takahashi, K. Segawa, Y. Ando, *Nature Phys.* **2012**, 8, 800.
[11] J. Shen, Y. Jung, A. S. Disa, F. J. Walker, C. H. Ahn, J. J. Cha, *Nano Lett.* **2014**, 14, 4183.
[12] Z. Li, S. Shao, N. Li, K. McCall, J. Wang, S. X. Zhang, *Nano Lett.* **2013**, 13, 5443.
[13] S. Guo, A. F. Fidler, K. He, D. Su, G. Chen, Q. Lin, J. M. Pietryga, V. I. Klimov, *J. Am. Chem. Soc.* **2015**, 137, 15074.
[14] M. V. Kovalenko, W. Heiss, E. V. Shevchenko, J.-S. Lee, H. Schwinghammer, A. P. Alivisatos, D. V. Talapin, *J. Am. Chem. Soc.* **2007**, 129, 11354.
[15] H. D. Jin, C.-H. Chang, *J. Mater. Chem.* **2011**, 21, 12218.
[16] K. Chang, T. P. Kaloni, H. Lin, A. Bedoya-Pinto, A. K. Pandeya, I. Kostanovskiy, K. Zhao, Y. Zhong, X. Hu, Q. K. Xue, *Adv. Mater.* **2019**, 31, 1804428.
[17] B. Dong, Z. Wang, N. T. Hung, A. R. Oganov, T. Yang, R. Saito, Z. Zhang, *Phys. Rev. Mater.* **2019**, 3, 013405.
[18] D. G. Stroppa, L. A. Montoro, A. Beltran, T. G. Conti, R. O. Da Silva, J. Andres, E. R. Leite, A. J. Ramirez, *Chem. Commun.* **2011**, 47, 3117.
[19] W. Wan, Y. Li, X. Ren, Y. Zhao, F. Gao, H. Zhao, *Nanomaterials* **2018**, 8, 112.
[20] F. Li, M. M. R. Moayed, F. Gerdes, S. Kull, E. Klein, R. Lesyuk, C. Klinke, *J. Mater. Chem. C* **2018**, 6, 9410.
[21] F. Li, M. M. Ramin Moayed, E. Klein, R. Lesyuk, C. Klinke, *J. Phys. Chem. Lett.* **2019**, 10, 993.
[22] S. Ghosh, L. Manna, *Chem. Rev.* **2018**, 118, 7804.
[23] M. Grzelczak, J. Pérez-Juste, P. Mulvaney, L. M. Liz-Marzán, *Chem. Soc. Rev.* **2008**, 37, 1783.
[24] F. Gerdes, E. Klein, S. Kull, M. M. R. Moayed, R. Lesyuk, C. Klinke, *Z. Phys. Chem.* **2018**, 232, 1267.
[25] X. Xia, Y. Xia, *Nano Lett.* **2012**, 12, 6038.

[26] F. Gerdes, C. Navío, B. H. Juárez, C. Klinke, *Nano Lett.* **2017**, 17, 4165.
[27] S. Kull, L. Heymann, A. B. Hungria, C. Klinke, *Chem. Mater.* **2019**, 31, 5646.
[28] C. Schliehe, B. H. Juarez, M. Pelletier, S. Jander, D. Greshnykh, M. Nagel, A. Meyer, S. Foerster, A. Kornowski, C. Klinke, *Science* **2010**, 329, 550.
[29] P. Giannozzi, S. Baroni, N. Bonini, M. Calandra, R. Car, C. Cavazzoni, D. Ceresoli, G. L. Chiarotti, M. Cococcioni, I. Dabo, *J. Phys-Condens. Mat.* **2009**, 21, 395502.
[30] X. Chen, D. Parker, D. J. Singh, *Sci. Rep.* **2013**, 3, 3168.
[31] E. Burstein, S. Perkowitz, M. Brodsky, *J. Phys. Colloq.* **1968**, 29, C4.
[32] L. Jones, H. Yang, T. J. Pennycook, M. S. Marshall, S. Van Aert, N. D. Browning, M. R. Castell, P. D. Nellist, *Adv. Structural & Chem. Imaging* **2015**, 1, 8.